\documentclass[sigconf]{acmart}

\makeatletter
\g@addto@macro\normalsize{%
  \abovedisplayskip 0pt plus2pt 
  \belowdisplayskip
  \abovedisplayskip
  \abovedisplayshortskip  0pt plus2pt%
  \belowdisplayshortskip  0pt plus0pt
}
\makeatother

\usepackage{balance}
  
\usepackage{booktabs} 
\usepackage{multirow}
\usepackage{subcaption}

\def\BibTeX{{\rm B\kern-.05em{\sc i\kern-.025em b}\kern-.08emT\kern-.1667em\lower.7ex\hbox{E}\kern-.125emX}}
    
\copyrightyear{2019}
\acmYear{2019}
\setcopyright{acmcopyright}
\acmConference[SIGIR '19] {42nd International ACM SIGIR Conference on Research and Development in Information Retrieval}{July 21--25, 2019}{Paris, France}
\acmBooktitle{42nd Int'l ACM SIGIR Conference on Research \& Development in Information Retrieval (SIGIR'19), July 21--25, 2019, Paris, France}
\acmPrice{15.00}
\acmDOI{10.1145/3331184.3331261}
\acmISBN{978-1-4503-6172-9/19/07}

\begin{document}

\title{Controlling Risk of Web Question Answering}

\author{Lixin Su, Jiafeng Guo, Yixing Fan, Yanyan Lan, and Xueqi Cheng}
\affiliation{
  \institution{
    CAS Key Lab of Network Data Science and Technology, Institute of Computing Technology, \\ Chinese Academy of Sciences, Beijing, China\\
    University of Chinese Academy of Sciences, Beijing, China\\} 
   }
\email{{sulixin,guojiafeng,fanyixing,lanyanyan,cxq}@ict.ac.cn}

\begin{abstract}
Web question answering (QA) has become an indispensable component in modern search systems, which can significantly improve users' search experience by providing a direct answer to users' information need. This could be achieved by applying machine reading comprehension (MRC) models over the retrieved passages to extract answers with respect to the search query. With the development of deep learning techniques, state-of-the-art MRC performances have been achieved by recent deep methods. However, existing studies on MRC seldom address the predictive uncertainty issue, i.e., how likely the prediction of an MRC model is wrong, 
leading to uncontrollable risks in real-world Web QA applications. In this work, we first conduct an in-depth investigation over the risk of Web QA. We then introduce a novel risk control framework, which consists of a qualify model for uncertainty estimation using the probe idea, and a decision model for selectively output.
For evaluation, we introduce risk-related metrics, rather than the traditional EM and F1 in MRC, for the evaluation of risk-aware Web QA. The empirical results over both the real-world Web QA dataset and the academic MRC benchmark collection demonstrate the effectiveness of our approach.

\end{abstract}

%
%
%

%

%
\maketitle

\section{Introduction}

It has been a long-term expectation that search systems can not only connect people to documents, but also connect people directly to information. One step towards this is to provide direct answers (e.g., facts, definitions or a short piece of text) to users' search queries. In this way, we can significantly improve users' search experience by saving their efforts on clicking and reading the result pages. This would be especially valuable in mobile search scenarios where browsing is quite difficult due to the limited screen size.

In modern search systems, the above target could be achieved by deploying a Web question answering (QA) component \cite{brill2002analysis,ferrucci2010building} upon the retrieval module: Given a search query and a set of top-ranked passages, the Web QA component determines whether the candidate set contains any direct answer and extracts the answer as the output if it exists. 
Without loss of generality, the Web QA component could be implemented by applying machine reading comprehension (MRC) models~\cite{chen2017reading, wang2017r, wang2017evidence,nishida2018retrieve} over the retrieved passages. In recent years, with the development of deep learning techniques, state-of-the-art MRC performances have been achieved by a variety of deep MRC models \cite{hu2018read+,devlin2018bert}.

However, directly applying existing MRC techniques for Web QA brings non-negligible risks in practice. Facing with open-domain long-tail queries and noisy search results, even the most advanced deep MRC models are prone to produce unreliable answers (i.e., overconfident incorrect answers \cite{dhamija2018reducing}). These unreliable answers may hurt users' search experience significantly, which we will discuss later in section \ref{sec:analysis}.  Therefore, for practical applications, it is expected that people can be aware of MRC models' confidence or uncertainty\footnote{We use confidence and (negative) uncertainty interchangeably in this paper.} \cite{gal2016uncertainty} on the predicted answers to enable risk control in Web QA. This calls for the investigation on the predictive uncertainty issue of MRC, a core research problem we want to tackle in this work. Unfortunately, most previous research on MRC has been focused on improving the model effectiveness \cite{wang2016machine,xiong2016dynamic,seo2016bidirectional,wang2017gated}. There has been little work addressing the predictive uncertainty issue of MRC. Note that there have been some studies tackling the unanswerable question (i.e., null answer) problem in MRC~\cite{rajpurkar2018know,hu2018read+,kundu2018nil}, which is different from the predictive uncertainty issue we discuss here.

In recent years, there has been increasing interest on the predictive uncertainty issue of deep learning models in the machine learning (ML) field. Existing work on predictive uncertainty can mainly be divided into three categories. The first class stems from Bayesian Neural Networks~\cite{brill2002analysis}, 
which aims to estimate the predictive uncertainty by introducing a prior distribution over the model parameters \cite{gal2016dropout,Malinin2018Predictive}. However, these methods are usually computationally expensive and the effectiveness largely depends on the correctness of the prior assumption. The second class~\cite{lakshminarayanan2017simple,geifman2018bias} borrows the ensemble idea, which tries to estimate the predictive uncertainty by model average based on a bag of models learned with different initialization or from different epochs. The third class~\cite{hendrycks2016baseline,mandelbaum2017distance,bahat2018confidence} takes the predictive uncertainty estimation as an additional task, where some heuristic strategies have been employed to learn another prediction model. However, most of these studies have been conducted in the Computer Vision (CV) field, with only a few in natural language processing (NLP), including semantic parsing and machine translation \cite{blatz2004confidence,dong2018confidence}. So far as we know, there has been little work tackling the predictive uncertainty issue of MRC.

In this paper, we propose to control the risk of Web QA by modeling the predictive uncertainty of deep MRC models\footnote{We focused on deep MRC models since they are most popular and advanced techniques for Web QA.}. We first take an in-depth analysis over the risk in the Web QA scenario and show its speciality as compared with the risks in other applications. Based on this analysis, we then introduce a novel risk control framework for Web QA with the \textit{selective classification} idea \cite{geifman2017selective}. Specifically, our framework consists of two major components, namely a \textit{qualify} model and a \textit{decision} model. The qualify model is used for predictive uncertainty estimation, which produces a confidence score for the prediction of an MRC model. Inspired by the probe idea introduced by Alain and Bengio \cite{alain2016understanding}, we design a novel and general \textit{PROBE-CNN} model to act as the qualify model based on the unified abstraction of deep MRC models. The decision model aims to learn a rejection region over the confidence score for selective output. By rejecting those low-confidence predictions of an MRC model, we can well control the risk of Web QA. Note that our framework is a post-processing framework which means it could be applied over almost any existing state-of-the-art deep MRC model.

For evaluation, traditional widely used metrics such as EM and F1 only focus on the effectiveness of an MRC model. In this work, we introduce risk-related metrics following the idea in \cite{geifman2017selective,geifman2018bias}, including \textit{coverage}, \textit{risk} and \textit{AURC}, for risk-aware Web QA evaluation. We conduct extensive experiments on two large scale publicly available datasets. One is a real-world WebQA dataset and the other is a widely adopted academic MRC benchmark collection~\cite{rajpurkar2016squad}. Two representative deep MRC models, i.e., BIDAF \cite{seo2016bidirectional} and BERT \cite{devlin2018bert} are employed, and  several state-of-the-art uncertainty estimation methods have been compared within our risk control framework. The experimental results show that our approach can outperform all the baseline methods in terms of all the evaluation metrics. Besides, we also provide detailed analysis to gain better understanding on our probe-based qualify model. 

The main contributions of this paper include:
\begin{itemize}
\item We introduce the risk control problem of Web QA which calls for addressing the predictive uncertainty issue of MRC models. So far as we know, this is the first work tackling such uncertainty issue of MRC models in Web search.
\item We propose a risk control framework for Web QA with a novel and general qualify model designed based on the probe idea and the unified abstraction of  deep MRC models.
\item We introduce risk-aware evaluation metrics for Web QA and conduct extensive experiments to demonstrate the effectiveness of our approach.
\end{itemize}

\section{Related Work}
In this section, we will briefly survey three related topics to our work, including Web QA, MRC, and model uncertainty. 

\subsection{Web QA}
Web QA is an important task across both NLP and information retrieval (IR) fields, which aims to answer users' questions using Web resources. Web QA is a type of open-domain QA in the sense that queries are usually from unconstrained categories and resources are typically unstructured Web documents. Without loss of generality, Web QA could be performed by a two-step process, i.e., relevant document retrieval and answer extraction. In this work, we focus on the answer extraction part. 

Early techniques and systems on Web QA have been largely driven by the TREC QA track~\cite{voorhees1999treca}. Most of these studies were based on shallow linguistic processing and complicated rules. For example, AskMSR~\cite{brill2002analysis} was a search-engine based QA system that relies on data redundancy to find short answers. Moldovan~\cite{moldovan2000structure} proposed window-based word scoring technique to rank potential answer pieces for Web QA.

Recently, researchers have released several large scale datasets, such as SearchQA\cite{dunn2017searchqa}, TriviaQA\cite{joshi2017triviaqa} and Quasar~\cite{dhingra2017quasar}, to accelerate the study and application of deep learning techniques for Web QA. Based on these datasets, Wang et al.~\cite{wang2017r} proposed to apply a deep ranker-reader model to extract answers from the top passages using reinforcement learning. Some other works~\cite{lin2018denoising,wang2017evidence} tried to employ deep MRC models to extract answers by either de-noising the data or re-ranking multiple candidate answers.
In this work, we focus on controling the risk of Web QA, instead of proposing another Web QA model.

\begin{figure*}[t]
\begin{subfigure}{0.45\textwidth}
\includegraphics[width=\linewidth]{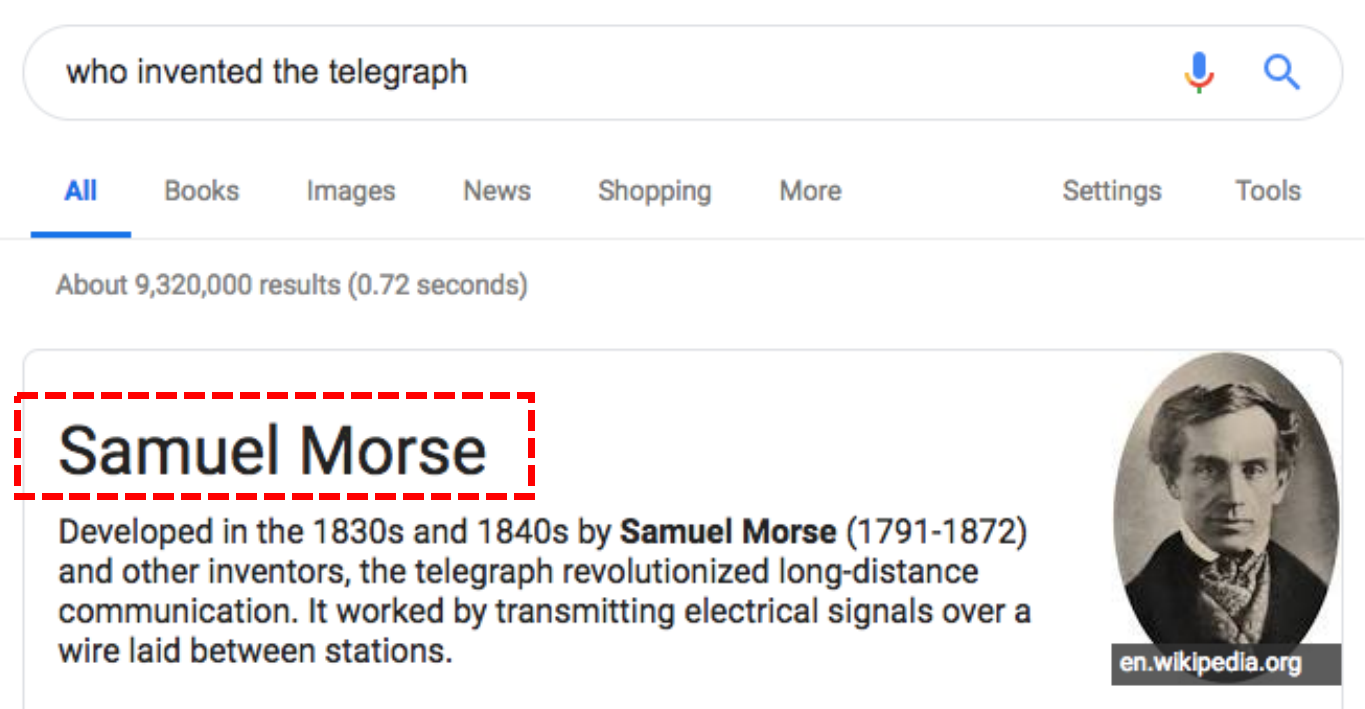}
\caption{A correct answer for the query ``\textit{who invented the telegraph}''.} 
\end{subfigure}
\hspace{0.02\textwidth}
\begin{subfigure}{0.45\textwidth}
\includegraphics[width=\linewidth]{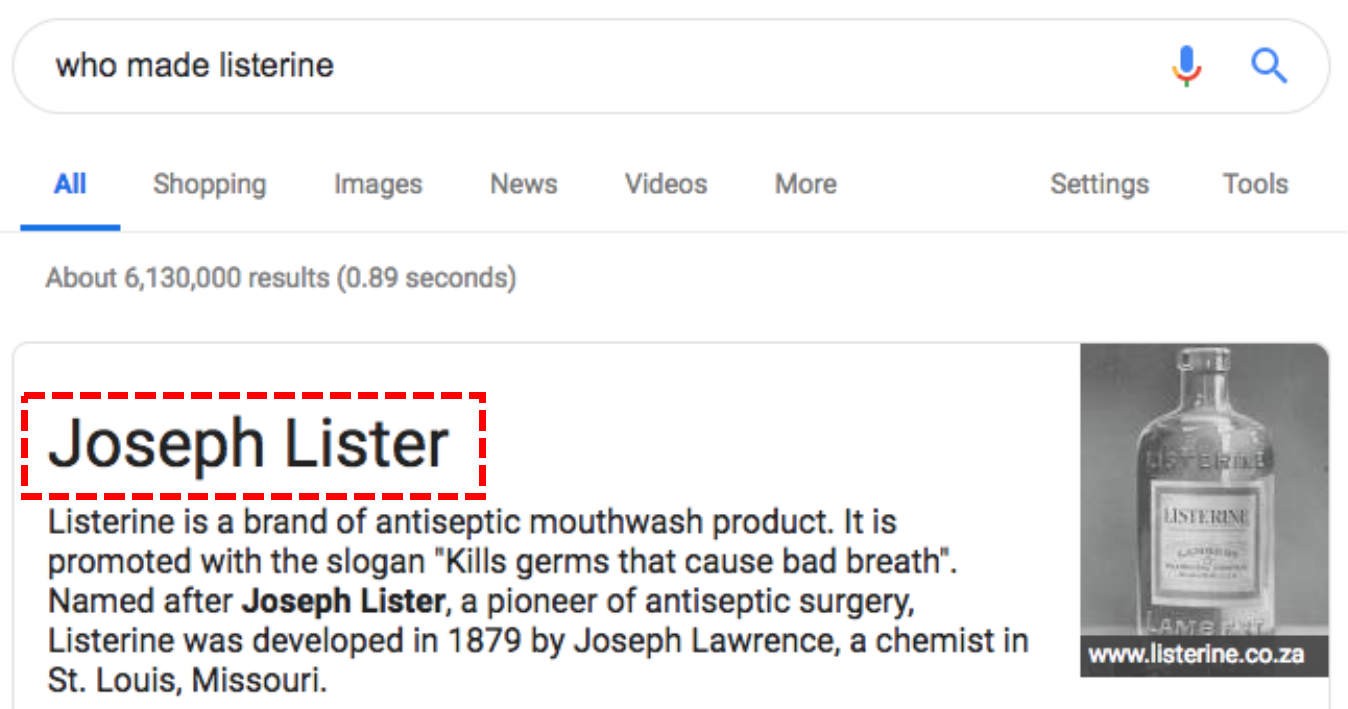}
\caption{A wrong answer for the query ``\textit{who made listerine}''.} 
\end{subfigure}
\caption{Direct answers from the Web QA component in modern search engines.} \label{fig:cases}
\end{figure*}

\subsection{Machine Reading Comprehension}
The MRC task is to predict the exact answer from a context passage for a given question \cite{chen2018neural}. According to the answer form, the MRC task can be further divided into four categories, namely cloze-style MRC~\cite{hermann2015teaching}, multi-choice MRC~\cite{richardson2013mctest,lai2017race}, span-prediction based  MRC~\cite{rajpurkar2016squad,rajpurkar2018know} and free-form MRC~\cite{nguyen2016ms}. Here, we focus on the span-prediction based MRC, which is more practical and prevalent, and also highly related to Web QA \cite{chen2017reading}.

In recent years, a large number of models have been proposed to address the span prediction task in MRC. In general, these models typically consist of two components, namely the reading component and the answering component. 
The reading component is used to capture the interactions between the question and the passage, and to collect the evidences from the passage. Researchers have introduced different types of architectures for the reading component, such as matching-LSTM~\cite{wang2016machine}, gated attention and self-attention mechanisms~\cite{wang2017evidence}, cross-layer interaction and hierarchical interaction \cite{hu2017mnemonic,wang2018multi,tay2018densely}. 
The answering component aims to extract the exact answer span from the passage based on the signals from the reading component. Most works \cite{wang2016machine} utilized pointer network~\cite{NIPS2015_5866} to predict the answer span in this component.

Early studies on MRC mainly focused on the questions whose answer definitely exists in the context passage. Recently, more attentions have been paid to the MRC task with unanswerable questions~\cite{rajpurkar2018know}. A common solution is to add a null position to the passage \cite{kundu2018nil,tan2018know}. Some work also introduced an additional model to detect those unanswerable questions~\cite{hu2018read+}.

Unlike most previous work which mainly focused on the effectiveness of the MRC model, our work focus on addressing its predictive uncertainty.

\subsection{Predictive Uncertainty in Deep Models}
Modeling predictive uncertainty~\cite{gal2016uncertainty} is critical for robust or safe AI. Recently, the uncertainty prediction problem has attracted a lot of interests, especially in the classification task. Without loss of generality, existing methods can be categorized into three classes.
The first class is based on the Bayesian Neural Network, which introduces a prior distribution over the model parameters \cite{mackay1992practical,gal2016dropout} to estimate the uncertainty. Their performance depends on the concrete form of approximation made due to computational constraints. Monte-Carlo Dropout is a kind of approximation using an ensemble of multiple stochastic forward passes and computing the spread of the results. However, the high computational cost still makes it intractable for complex deep neural networks.
The second class uses the ensemble idea to estimate the predictive uncertainty, with the assumption that multiple predictions on the same instance can provide information about the uncertainty. For example, 
Geifman et al.~\cite{geifman2018bias} proposed to ensemble model information from different training epochs for the uncertainty estimation.
The third class takes uncertainty estimation as an additional prediction problem, and employs some heuristic function for estimation. In \cite{hendrycks2016baseline}, the simple \textit{max} function was applied to the model's last output to produce the uncertainty score. More complex functions such as local density based functions~\cite{mandelbaum2017distance} have also been utilized for estimation.

Most previous work on predictive uncertainty estimation has been proposed in the ML community and mainly applied on CV tasks. To the best of our knowledge, this is the first work to address the predictive uncertainty issue of MRC for the Web QA task.

\section{Analysis on the Risk of Web QA} \label{sec:analysis}

Web QA has become a key feature in modern search engines. As shown in Figure~\ref{fig:cases}(a), given a search query "who invented the telegraph", beyond the traditional \textit{ten blue links}, a direct answer "Samuel Morse" is provided through Web QA. In this way, users' information need could be satisfied without browsing any returned Web page. However, Web QA might also face unexpected risks in practice. For example, as shown in Figure~\ref{fig:cases}(b), the returned direct answer for the query "who made listerine" is incorrect since its true developer is "Joseph Lawrence" as shown in the corresponding passage. Such incorrect answers might mislead users, and even become harmful if this happens for search queries seeking for medical or legal suggestions. In the following, we try to study two research questions to gain a better understanding of the risk in the Web QA scenario.

\begin{table}[]
\caption{ Categorization of the output in Web QA.}
\label{table:cost}
\centering
\begin{tabular}{|c|c|c|c|}
\hline
\multirow{2}{*}{} & \multicolumn{3}{c|}{Model Prediction}                   \\ \cline{2-4} 
                  & \multicolumn{2}{c|}{Direct Answer(D)} & Null Answer(N)  \\ \hline
Answerable(A)     & AD$^+$            & AD$^-$            & AN              \\ \hline
Unanswerable(U)   & \multicolumn{2}{c|}{UD}               & UN              \\ \hline
\end{tabular}

\end{table}

Firstly, what are the true risk of Web QA? As shown in Table~\ref{table:cost}, the input queries\footnote{Note that in practice there would be a triggering component (i.e., an intent classifier) which determines whether to trigger the Web QA component, i.e., whether a query is likely to have a short answer. However, this is out of the scope of this paper and here we refer to input queries as those have passed such triggering component.} in Web QA could be classified into two categories, i.e., answerable (A) and unanswerable (U). Answerable queries refer to the queries whose answer exists in the top returned results, while unanswerable queries are the opposite. Meanwhile, considering the model prediction, a QA model may output a direct answer (D), either correct (+) or not (-), or a null answer (N) for a given query. Therefore, the output of Web QA must fall into one of the five folds, where AD$^+$/AD$^-$ denotes that the model predicts an correct/incorrect answer for an answerable query respectively, AN/UN denotes that the model predicts a null answer for an answerable/unanswerable query respectively, and UD denotes that the model predicts a direct answer for an unanswerable query. 
As we can see, there is obviously no risk in the AD$^+$ and UN folds. The true risks of Web QA only come from the AD$^-$ and UD folds, where the model predictions are incorrect with respect to the ground truth. It is noteworthy that there is no risk in the AN fold either, even though the model's prediction is wrong. This is because users' search experience may not be influenced when the model provides a null answer. Such a unique risk structure makes Web QA  different from those CV or NLP tasks (which do not have null answer predictions) where risks have been tackled before.

Secondly, what causes the risk of Web QA? 
In the Web QA scenario, when we rely on MRC techniques for answer prediction, the risk happens when the MRC model cannot provide reliable predictions but people are not aware of it. There are two major reasons related to this problem. On one hand, although MRC models can achieve human-like performance on some close-world dataset \cite{rajpurkar2016squad}, it will encounter significant challenges in Web QA due to its open-domain nature. Facing with long-tail queries and noisy search results, even the most advanced deep MRC models cannot generalize well over many previously unseen QA patterns. On the other hand, recent studies have revealed that deep models are poor at uncertainty qualification \cite{guo2017on} and tend to produce overconfident predictions \cite{dhamija2018reducing}. In other words, although deep MRC models could provide a probability on its answer, that probability cannot well reflect its confidence on the prediction. In summary, deep MRC models are prone to produce overconfident but incorrect answers, leading to the risk of Web QA. 

Based on the above analysis, we can find that Web QA is a special problem with its unique risk structure. The risk of Web QA comes from people's unawareness of the model uncertainty on its predictions. Therefore, in the following, we try to control the risk of Web QA by modeling the predictive uncertainty of deep MRC models.

\section{Risk Control Framework} \label{sec:frame}
\begin{table}[t]
 \caption{Basic notations used in this paper.}
 \label{table:notation}
\begin{tabular}{ll}
\hline
Meaning                                                       & Notation      \\ \hline
Query space                                                   & $\mathcal{Q}$ \\
Passage space                                                 & $\mathcal{P}$ \\
Answer space                                                  & $\mathcal{A}$ \\
Query representation matrix  in layer $t$                     & $Q^{(t)}$       \\
Passage representation matrix in layer $t$                    & $P^{(t)}$       \\
Start position vector of the answer span & $\vec{s}$   \\
End position vector of the answer span  & $\vec{e}$   \\ \hline
\end{tabular}

\end{table}

In this section, we describe the risk control framework for Web QA. Note that we do not aim to propose a specific MRC model that can qualify its uncertainty better than existing methods, but rather design a general framework that could work for a large variety of MRC models. In this way, our framework could be easily integrated with the practical Web QA system and do not restrict any future upgrade of the applied MRC models. Towards this purpose, we adopt the third class modeling methodology on predictive uncertainty, which takes the uncertainty estimation as an additional prediction problem, and make the framework a post-processing one so that it could be applied over any existing state-of-the-art deep MRC models. A key difference of our framework from previous uncertainty estimation work is that we do not simply employ some heuristic estimation functions, but rather introduce a general probe-based uncertainty prediction model specifically designed for modern deep MRC models.

Some basic notations frequently used in this paper are listed in Table~\ref{table:notation}. Overall, our framework takes an MRC model as the input, and learns two new models, namely the qualify model and the decision model, to control the risk. 
The formal definition of each model is as follows. 

\textbf{The MRC model $f$} is a function which predicts an answer $a$ based on a passage $p$ with respect to a search query $q$, denoted as $f:(\mathcal{Q},\mathcal{P}) \to \mathcal{A}$.
Note here we only consider the MRC model $f$ which takes a single passage for direct answer prediction for simplicity (which is also a typical case in practical Web QA where only the top relevant passage is used due to the efficiency concern). However, our framework is not limited to that, but could well adapt to answer extraction from multiple retrieved passages.

\textbf{The Qualify model $g$} is core in our framework, which aims to estimate the predictive uncertainty of the MRC model. Specifically, given an MRC model $f$ and a specific query-passage instance ($q$,$p$), the qualify model $g$ outputs a confidence score for $f$'s prediction, denoted as $g(q,p|f) \in [0,1]$. Here, the confidence score represents the likelihood that $f$'s prediction is correct. The detailed implementation of the qualify model $g$ will be described in Section~\ref{implementation}.

\textbf{The Decision model $h$} is used to make the final decision whether we shall abandon $f$'s prediction by defining a rejection region over the confidence score,
\begin{equation}
h(g)\triangleq
\begin{cases}
1 & \text{if } g(q,p|f) \geq \theta  \\
0 & \text{if } g(q,p|f)  < \theta,
\end{cases}
\label{eq:h}
\end{equation}
where $\theta\in [0,1]$ denotes the confidence threshold parameter for decision. When $h(g)=1$, we choose to trust/output $f$'s prediction, otherwise not.

\textbf{Learning of the framework} could be derived as follows. Given an MRC model $f$, we can write down the \textit{risk} as
\begin{equation}
  R(f|P) = E_{Pr(q,p,a)}[\ell(f(q,p), a)],
\end{equation}
where $\ell: \mathcal{A}\times\mathcal{A}\to \mathbb{R}^+$ is a given loss function, and $Pr(q,p,a)$ is an unknown data distribution over $\mathcal{Q}\times\mathcal{P}\times\mathcal{A}$. Given a labeling set $S_m=\{(q_i,p_i,a_i)\}_{i=1}^m \in (\mathcal{Q}\times\mathcal{P}\times\mathcal{A})$, the \textit{empirical risk} is as follows:
\begin{equation}
    \hat{r}(f,h,g|S_m) \triangleq \frac{\sum_{i=1}^{m}\ell(f(q_i,p_i),a_i)}{m}.
\end{equation}
Based on the analysis in Section~\ref{sec:analysis}, the loss function in Web QA could be defined as follows
\begin{equation}
\ell(f(q,p), a) \triangleq
\begin{cases}
0 & \text{$f(q,p) \in AD^+ \cup AN \cup UN $ } \\
1 & \text{$f(q,p) \in AD^- \cup UD$}. \\
\end{cases}
\end{equation}

In traditional uncertainty estimation, an \textit{optimal} qualify model $g$ (for $f$) should reflect true loss monotonicity in the sense that the confidence score should be higher for the instance with lower loss. However, due to the special risk structure in Web QA, this is not always the case. Specifically, although there is no loss in the AN set (i.e., null answer for answerable query) as shown in Table~\ref{table:cost} , we do not expect a high confidence in this area since the predictions are actually wrong. Moreover, we could omit the estimation for instances in the UN (i.e., null answer for unanswerable query) and AN set, since there would be no gain even if we predict a confidence score for those instances.
Therefore, we only need to require the monotonicity between instance $f(q_i,p_i)\in AD^+$ and $f(q_j,p_j)\in AD^-\cup UD$,
\begin{equation}
g(q_i,p_i|f) \geq g(q_j,p_j|f). 
\end{equation}
Based on this target, it turns out we could optimize the following pairwise objective function in order to learn the qualify model $g$,
\begin{equation}
   loss = max(0, 1 - g(q_i,p_i|f) + g(q_j,p_j|f) ),
\label{eq:pair}
\end{equation}
where $f(q_i,p_i)\in AD^+$ and $f(q_j,p_j)\in AD^-\cup UD$.

With the learned qualify model $g$, we can use the decision model $h$ to selectively output $f$'s prediction and obtain the following \textit{empirical selective risk}
\begin{equation}
    \hat{r}(f,h,g|S_m) \triangleq \frac{\sum_{i=1}^{m}\ell(f(q_i,p_i),a_i)h(g)} {\sum_{i=1}^{m}h(g)}.
\label{eq:em_risk}
\end{equation}
Note here $\hat{r}(f,h,g|S_m)\in[0,1]$. Finally, to decide the confidence threshold $\theta$ in $h$ in practice, we only need to setup a desired risk level and find the $\theta$ that satisfies the risk level over $S_m$ (which is not necessarily the training set).

\begin{figure*}[t]
\centering
\includegraphics[scale=0.3]{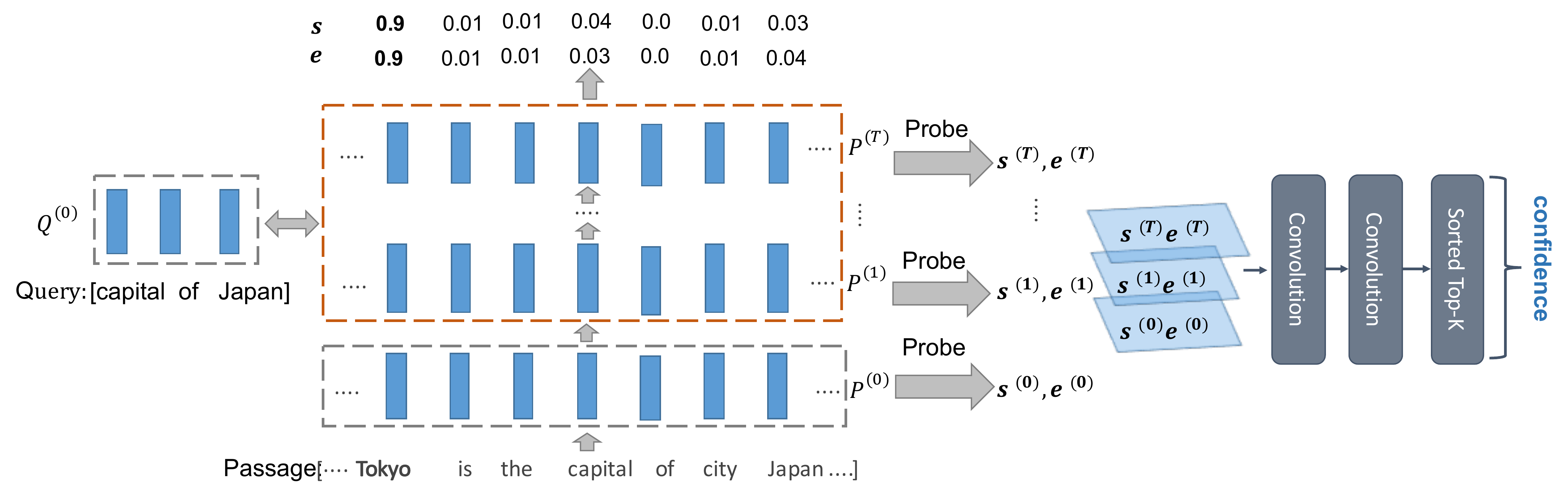}
\caption{The architecture of the deep qualify model.}
\label{fig:framework}
\end{figure*}

\section{Quality Model Implementation} \label{implementation}

In this section, we focus on describing the core component in our risk control framework, i.e., the qualify model. As mentioned above, our goal is to construct a general risk control framework that could be applied over a large variety of deep MRC models. This in turn requires that the qualify model could adapt to different MRC models. In this paper, we mainly focus on deep MRC models designed for answer span prediction, which are prevalent and state-of-the-art techniques in Web QA. We will leave other MRC models, e.g., traditional models or generative models, for the future work. 

Our main design for a general qualify model is inspired by the probe idea \cite{alain2016understanding}, which attempts to use linear-classifier probes to understand neural network models. In this work, we try to probe deep MRC models to estimate predictive uncertainty in a unified way. Specifically, we first try to abstract modern deep MRC models into a unified view. Based on this unified view, we show how to insert probes into MRC models to extract useful signals for uncertainty estimation. Finally, based on the learned probes, we show how to build an estimation model to produce the confidence score via supervised learning.

\subsection{A Unified View for Deep MRC Models}

Many studies on MRC formulate the QA task as an answer span prediction problem. Given a (question) query\footnote{To avoid confusion on the terminology, we view a question as a query so that we can use the term query for consistency.} and a passage, an MRC model extract an answer (i.e., a segment of text in the passage) if and only if the passage contains an answer. When the passage does not contain any answer, the model returns \textit{null}. 

Formally, the passage and the query are described as a sequence of word tokens, denoted as $p=\{w^p_i\}^{l_p}_{i=1}$ and $q=\{w^q_j\}^{l_q}_{j=1}$ respectively, where $l_p$ is the passage length and $l_q$ is the query length. Then the extracted answer could be denoted as $a=\{w^p_i\}^{e}_{i=s}$, where $s$ and $e$ denotes the start and end position of the answer span respectively. If there is no answer, $s$ and $e$ would point to a null position. In this way, the learning objective of MRC becomes to learn a model $f$ to maximize the log-likelihood
\begin{equation}
    \log Pr(a|q,p) = \log Pr(s,e|q,p).
\end{equation}
Then in prediction,
\begin{equation}
\hat{s}, \hat{e} = \text{argmax}_{s, e \in \mathcal{L}(p)} Pr(\hat{s}, \hat{e}|q, p),
\label{eq:decode}
\end{equation}
where $\mathcal{L}(p)$ denotes all the possible start-end position pairs in $p$. 

With the development of deep learning techniques, most work on MRC implements $f$ by deep neural networks and continuously refreshes the state-of-the-art performance. Different complicated architectures have been designed for deep MRC models, typically including components like embedding, sequential encoding, convolution, self-attention, co-attention and so on. However, to design a qualify model that can be applied on these different deep MRC models, we need a unified view over these existing models.

In this work, we find that although existing deep MRC models on answer span prediction have quite different architectures, they all can be abstracted as the following process shown in Figure \ref{fig:framework}: 
\begin{enumerate}
\item[1.] Convert the text sequence $q$ and $p$ into their initial embedding representations $Q^{(0)} \in R^{l_q \times h^{(0)}_d}$ and $P^{(0)} \in R^{l_p \times h^{(0)}_d}$, where $h^{(0)}_d$ denotes the embedding dimension in the first layer; 
\item[2.] Distill the passage representation with a series of steps \\ $P^{(1)},P^{(1)},\dots,P^{(T)}$ through a variety of complicated interactions (e.g., self-attention, inter-attention and so on); 
\item[3.] Compute the start and end position vector $\vec{s}$ and $\vec{e}$ based on the final distilled representation $P^{(T)}$ and decode the optimal span $[s,e]$ according to Equation~(\ref{eq:decode}).
\end{enumerate}

Based on the above unified view, we find that deep MRC models in essence are about distilling useful representations of the passage for distinguishing the start and end position of the answer span. Such a common process inspires us to look into the intermediate distilled representations to extract useful signals for predictive uncertainty estimation, leading to the following probe-based method.

\subsection{Probing MRC Models}
The original probe idea proposed by Alain and Bengio \cite{alain2016understanding} was to monitor the features at every layer of a model and measure how suitable they are for classification. In this work, we borrow the idea to investigate each distilled passage representation in a deep MRC model and measure how likely they are for distinguishing the start and end position of the answer span.

Specifically, given the $t$-th layer passage representation $P^{(t)}$, for each word position, we apply a linear layer over its latent representation to obtain a start and end score respectively, and then normalize the scores from all the positions through softmax to obtain the final start and end vector respectively,
\begin{equation}
	\vec{s}^{(t)} =  softmax(P^{(t)} \mathbf{v}_s^{(t)}),
\label{eq:probe1}
\end{equation}
\begin{equation}
	\vec{e}^{(t)} =  softmax(P^{(t)} \mathbf{v}_e^{(t)}).
\label{eq:probe2}
\end{equation}
where $\mathbf{v}_s^{(t)}$ and $\mathbf{v}_e^{(t)}$ are probe parameters for the $t$-th layer. Note here we use a linear probe due to its convexity property \cite{alain2016understanding}. We can avoid the issue of local minima since training a linear probe using softmax cross-entropy is a convex problem.
Each value in $\vec{s}^{(t)}$ and $\vec{e}^{(t)}$ actually denotes the possibility of a position to be the start and end position of the answer span, respectively. Therefore, the probe parameters can be learned so as to minimize the cross-entropy loss which is usually used for the last layer of the MRC model, 
\begin{equation}
	loss_{probe}^{(t)} = -log(\vec{s}_{s}^{t}) -  log(\vec{e}_{e}^{(t)}).
\end{equation}
Note that the probes do not affect the learning and prediction of the deep MRC model. They only measure the level of discrimination of the passage representation at a given layer.

With the learned probes, for each query-passage instance and a given MRC model, we can get a series of signals $\{(\vec{s}^{(t)},\vec{e}^{(t)})\}_{t=1}^{T}$. The signals are essential for the following uncertainty estimation.

%

\subsection{Predictive Uncertainty Estimation}
For a given MRC model and a query-passage instance, the predictive uncertainty estimation takes the probed signals above as the input to produce a confidence score within the range $[0,1]$. Here we build a deep model, namely PROBE-CNN, to achieve this purpose as shown in Figure \ref{fig:framework}. Specifically, we concatenate the start and end vectors in each layer into a feature matrix. We stack these feature matrices layer by layer from bottom to top. We then apply two convolution layers and a sorted top-k layer over the stacked representations. Finally, we feed the top-k signals into a fully connected layer to produce the final confidence score.

The underlying design idea of the above PROBE-CNN model is as follows. We view the probed signals from each layer of the MRC model as a ``X-ray'' picture of the distilled passage representation in that layer. Each X-ray picture indicates how likely that particular distilled passage representation is for distinguishing the start and end position of the answer span. With these X-ray pictures in hand, we try to contrast and summarize them to diagnose the predictive uncertainty, which is achieved by the convolution process. We plug in a sorted top-k layer since we consider the uncertainty more related to the signal distribution rather than position patterns. In our expectation, if the probed signals are unstable between layers or  distributed uniformly, there will be large uncertainty in prediction. We will verify this in our experiments.

The deep qualify model is learned on a validation set for generalization. Specifically, we first train a MRC model together with the probes on the training set. We then apply the learned MRC model and the probes on the validation set to obtain the prediction labels shown in Table~\ref{table:cost}  as well as the probed signals. Finally, we train the qualify model based on these labels and probed signals according to the objective function as shown in Equation~(\ref{eq:pair}).

\section{Experiments}
In this section, we conduct experiments to demonstrate the effectiveness of our risk control framework. We first introduce the the experimental settings, including datasets, basic MRC models, baseline methods and evaluation metrics. We then present the major comparison results on the benchmark collections. Furthermore, we also provide detailed analysis over the core component in our framework, i.e., the deep qualify model.

\begin{table}[]
\caption{Dataset statistics. \# denotes number, |len$_p$| denotes average query length, |len$_q$| denotes the average passage length, and null\% denotes the proportion of unanswerable instances.}
\label{table:data}
\small
\begin{tabular}{ccccccc}
\toprule \hline
                            & \#train                     & \#dev                      & \#test                     & |len$_p$| & |len$_q$| & null\%\\ \hline
\multicolumn{1}{l}{SogouRC} & \multicolumn{1}{l}{200,000} & \multicolumn{1}{l}{49,566} & \multicolumn{1}{l}{49,863} & 80.47     & 9.66      & 60\%   \\
SQuAD 1.0                   & 72,599                       & 15,000                      & 10,570                      & 116.98    & 10.2      & 0      \\
SQuAD 2.0                   & 115,319                     & 15,000                     & 11,873                     & 127.90    & 10.02     & 35\%  \\
\hline \bottomrule
\end{tabular}

\end{table}

\begin{table*}[t]
\caption{Main results of two MRC models under our framework. The number in the parentheses denotes the relative improvement of the model against PROBA.}
\label{table:main}
\centering
\begin{tabular}{cccccccc}
\toprule
\hline
\multicolumn{2}{c}{\multirow{2}{*}{}} & \multicolumn{3}{c}{SogouRC} & \multicolumn{3}{c}{SQuAD} \\ \hline
\multicolumn{2}{c}{}                  & AURC     & ROC     & AP     & AURC     & ROC    & AP    \\ \hline
\multirow{4}{*}{BIDAF}   & PROBA      &   21.89&73.44&61.62         &     32.68&65.65&58.43         \\
                         & AES        &   21.75(0.6\%) & 73.51(0.1\%) & 60.53(1.8\%) & 32.47(0.6\%) & 66.46(1.2\%) & 59.13(1.2\%)   \\
                         & ENS        &  21.18(3.2\%) & 74.71(1.7\%) & 62.53(1.5\%) & 31.35(4.1\%) & 67.02(2.1\%) & 60.89(4.2\%) \\
                         & PROBE-CNN       & \textbf{19.53}(10.8\%) & \textbf{76.39}(4.0\%) & \textbf{63.8}(3.5\%) & \textbf{31.09}(4.9\%) & \textbf{67.41}(2.7\%) & \textbf{60.98}(4.4\%)  \\ \hline 
                         & oracle     & 7.9&100.0&100.0    &      11.01&100.0&100.0             \\ \hline \midrule
\multirow{4}{*}{BERT}    & PROBA      &          15.28&74.21&52.19      &       27.02&64.32&51.83          \\
                         & AES        &          14.95(2.2\%) & 74.36(0.2\%) & 52.35(0.3\%) & 26.4(2.3\%) & 66.19(2.9\%) & 52.25(0.8\%)        \\
                         & ENS        &          14.58(4.6\%) & 74.94(1.0\%) & 53.3(2.1\%) & 23.99(11.2\%) & 68.58(6.6\%) & 55.22(6.5\%)         \\
                         & PROBE-CNN       &          \textbf{14.16}(7.3\%) & \textbf{75.46}(1.7\%) & \textbf{53.8}(3.1\%) & \textbf{23.41}(13.4\%) & \textbf{69.55}(8.1\%) & \textbf{56.47}(9.0\%)           \\ \hline
                         & oracle     &          5.23&100.0&100.0       & 7.92&100.0&100.0         \\ \hline 
                                \bottomrule
\end{tabular}

\end{table*}

\subsection{Datasets}
We conduct experiments on two publicly available datasets. The detailed statistics of these two datasets are shown in Table~\ref{table:data}.

\textbf{SogouRC}~\footnote{http://huodong.sogou.com/cips-sogou\_qa/} is a large scale Web QA dataset released by the Chinese commercial search engine Sogou. It includes $30,000$ queries selected from search logs that can be satisfied by short answers, and the corresponding top ranked passages from the search result. After pre-processing, we obtain $299,429$ query-passage pairs, among which around $40\%$ are answerable instances, and the rest are unanswerable instances.

\textbf{SQuAD}~\cite{rajpurkar2018know} is a large scale MRC datasets with two versions. SQuAD 1.0 only contains answerable questions. SQuAD 2.0 combines the SQuAD 1.0 dataset with unanswerable questions. Although SQuADs are not typical Web QA datasets from search engines, here we use them for experiments due to the following reasons: 1) SQuAD 2.0 contains both answerable and unanswerable queries which are very similar to the Web QA scenario; 2) SQuADs have been widely adopted in MRC related research and our work addresses the predictive uncertainty of MRC.

\subsection{Basic MRC models}

As described in the section~\ref{sec:frame}, our risk control framework can be applied to a variety of existing deep MRC models. 
Here we take two representative deep MRC models BIDAF~\cite{seo2016bidirectional} and BERT~\cite{devlin2018bert} as the basic MRC model in our framework. 

\textbf{BIDAF} is a RNN-based MRC model. It first maps each word in the passage and query to the embedding space by combining character-level and word-level embeddings. LSTM layers with attention are then applied to collect contextual information, update the passage representation with respect to the query representation, and distill the representation of the passage. Finally, two vectors indicating the distribution of the start and the end index are derived from the final passage representation. 
Note that BIDAF model was originally designed only for answerable queries. To handle unanswerable queries in Web QA, we use the updated BIDAF model \cite{levy2017zero-shot} that can handle null answers~\footnote{We use the implementation: https://bit.ly/2rDHBgY}.

\textbf{BERT}~\footnote{We use the implementation: https://bit.ly/2B4TjGi} is a universal language representation model and can be used for many NLP tasks. The main structure of BERT  is a multi-layer transformer encoder. Specifically, for MRC, it first maps each word in the query and the passage its word embedding, position embedding and segment embedding. It then interacts the query and the passage to distill the passage representation. Based on the final-layer representation of the passage, the start and end vectors can be derived by a single linear layer to locate the answer in the passage. For recognizing unanswerable queries, BERT places a special token in the passage to indicate the null answer.

\subsection{Baselines}
Since the core component in our risk control framework is the qualify model, here we consider several existing predictive uncertainty estimation methods as our major baselines, including manually designed functions and ensemble-based methods.

\textbf{PROBA}~\cite{hendrycks2016baseline} simply employs a heuristic \textit{max} function on the normalized output probability of a model as the uncertainty estimation. Specifically, for deep MRC models, PROBA takes the max probability of spans in the passage 
$$max(softmax(\vec{s}) \otimes softmax(\vec{e}))$$
as the confidence score. This is the most intuitive method for estimating the confidence in MRC.

\textbf{AES}~\cite{geifman2018bias} makes use of the ensemble idea which averages model predictions from different epochs to estimate the predictive uncertainty. To adapt this method to deep MRC models, we train basic MRC models and save the model snapshot in each learning epoch. Note that it takes 20 epochs for BIDAF and 3 epochs for BERT to reach convergence. We then average the predicted start and end vectors from different model snapshots, and compute max probability as PROBA to obtain the final confidence score.

\textbf{ENS}~\cite{lakshminarayanan2017simple} trains multiple models from different initialization to estimate the predictive uncertainty of deep models on the image classification task. To adapt this model to the Web QA scenario, we train deep MRC models multiple times with different initialization. We then average the predicted start and end vectors, and compute the max span probability as the confidence score.
We tried different initialization numbers and find that with 3 randomly initialized models we can achieve good effectiveness-efficiency trade-off. Further increase on the initialization number brings little gain with much larger computational cost.

\subsection{Evaluation Metrics}
For evaluation, we introduce some widely adopted metrics in previous work on predictive uncertainty estimation \cite{lakshminarayanan2017simple,geifman2018bias} for risk-aware Web QA evaluation. 

Firstly, the performance of the risk control framework could be qualified using \textit{risk} and \textit{coverage}. 
The coverage is the probability mass of the non-rejected region of the confidence score, 
\begin{equation}
\hat\phi(f|S_m)\triangleq \frac{1}{m}\sum_{i=1}^{m}h(g).
\end{equation}
The risk is defined as the empirical loss in the non-rejected region
$$ \hat{r}(f,h,g|S_m) \triangleq \frac{\frac{1}{m}\sum_{i=1}^{m}\ell(f(q_i,p_i),a_i)h(g)} {\hat\phi(f|S_m)},$$
which is the same as the empirical selective risk defined in Equation~(\ref{eq:em_risk}). 
These two measures can be empirically evaluated over any finite labeled set $S_m$
(not necessarily the training set) in a straightforward manner whenever we choose a specific threshold parameter $\theta$ in the decision model. Moreover, these two metrics are actually trade-off to each other in the sense that if we increase $\theta$, we will often observe worse (i.e., lower) coverage with better (i.e., lower) risk. Therefore, we can draw the \textit{risk-coverage curve} (RC-curve) to show the trade-off relation. 

Now we introduce the metric \textit{area under the (empirical) RC curve} (AURC) as an overall performance measure for the risk control framework that is free of the specific selection of the threshold parameter $\theta$.
\begin{equation}
AURC(f,g|V_n) = \frac{1}{n}\sum_{h \in \mathcal{H}}\hat{r}(f,h,g|S_m).
\end{equation}
The smaller the AURC, the better the risk control framework is. Note that we can not only use the AURC to evaluate different risk control framework based on the same basic MRC model, but also compare them across different MRC models since AURC also reflects the effectiveness of the MRC model itself (when coverage equals 1).

Besides, we can also view the risk control framework as a filter which filters out the incorrect predictions of an MRC model. In this way, the risk control framework can be view as a binary classifier and we can measure its performance via \textit{the area under the ROC curve} (ROC) and \textit{the average precision} (AP) metrics.
The difference of these two metrics from the AURC is that they are independent of the MRC model. They only measure the performance of the qualify model. Opposite to the AURC, the larger ROC or AP, the better the qualify model is.

\subsection{Main Comparison}

\begin{figure}[t]
\centering
\includegraphics[scale=0.42]{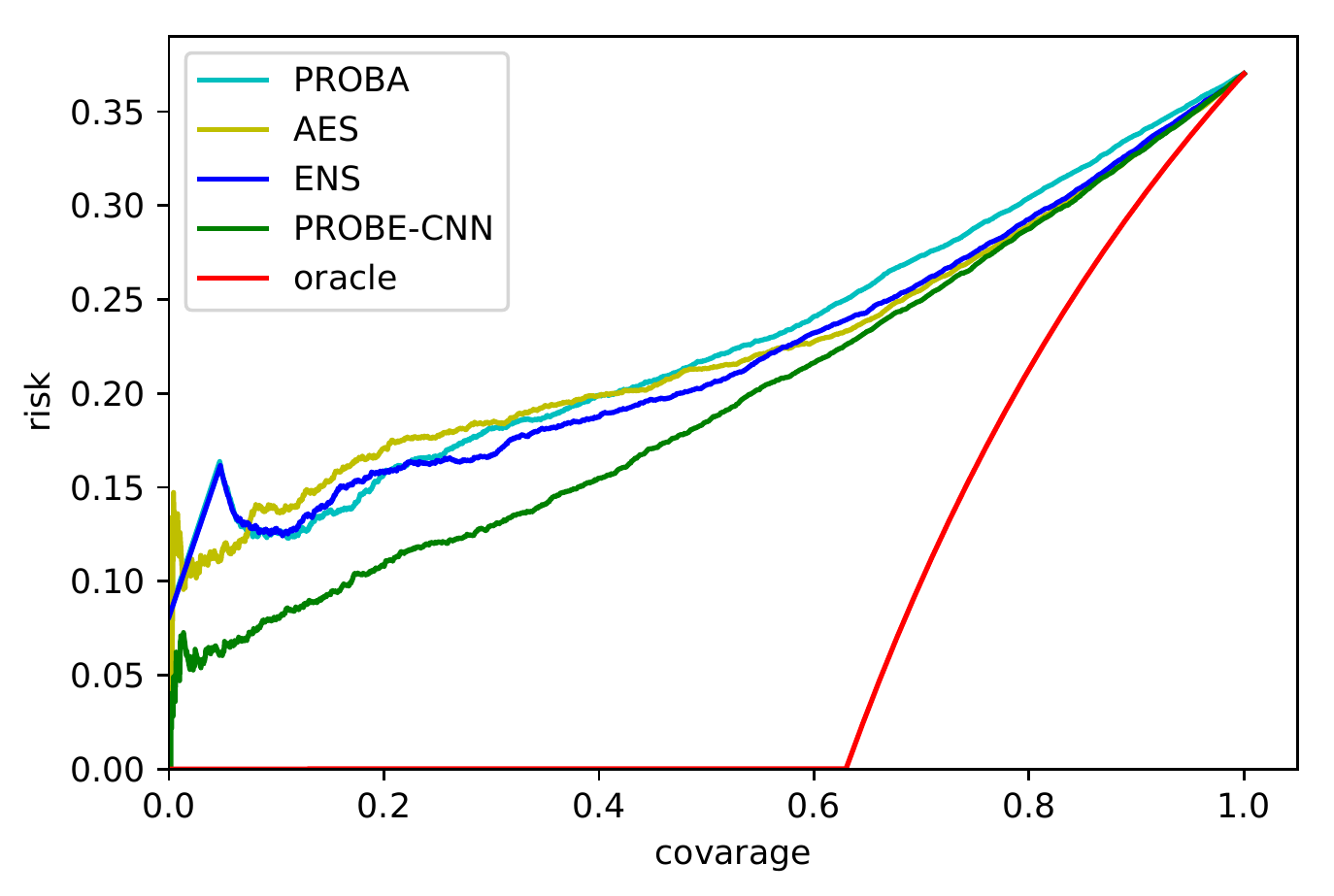}
\caption{Risk-coverage curve of BIDAF on SogouRC.}
\label{fig:rc_curve}
\end{figure}

We present the main comparison results of the risk control framework using different qualify models in Table~\ref{table:main}. In the table, we also show the \textit{oracle} result, which refers to a perfect risk control framework that can correctly identify instances in AD$^-$ and UD from those in AD$^+$. In this way, the ROC and AP of the oracle is always 1, while the AURC of the oracle can reflect the inherent ability of the MRC model as well as the dataset characteristic.

From the results, we have the following observations: (1) BERT can always obtain lower AURC score than BIDAF on the same dataset in the oracle mode, which indicates higher effectiveness of BERT over BIDAF on answer extraction. (2) Using the same basic MRC model, SogouRC always obtains lower AURC score than SQuAD 2.0 in the oracle mode, which indicates the SogouRC dataset is easier than the SQuAD 2.0. This is somehow counter-intuitive to us since the SogouRC dataset is an open domain Web QA dataset collected from the real-world search engine. However, after careful investigation, we find that SQuAD 2.0 is actually more difficult than SogouRC. The reason is that manually designed unanswerable queries in SQuAD 2.0 are much more difficult to identify (since the queries are often highly related to the passages) than those real-world unanswerable queries (which are unanswerable often due to the lack of related passages). (3) The ensemble-based methods, including AES and ENS, obtain better results than the manually designed PROBA method in terms of all the evaluation metrics. The results show that by using the average of multiple model variants (either from different epochs or from different initializations), we can obtain better predictive uncertainty estimation. However, ensemble-based methods usually require large computational cost which may not well fit the online requirement. (4) Our proposed PROBE-CNN model outperformd all baseline methods consistently in terms of all the evaluation metrics. For example, on the SogouRC dataset, the relative improvement of our method over the best-performing baseline ENS is about 7.7\% in terms of AURC. It is worth to note that our method is orthogonal to the ensemble-based method, in the sense that we can further include ensemble-based idea into our framework to improve the risk control performance. We will leave this as our future work.

\begin{figure}[t]
\centering
\includegraphics[scale=0.3]{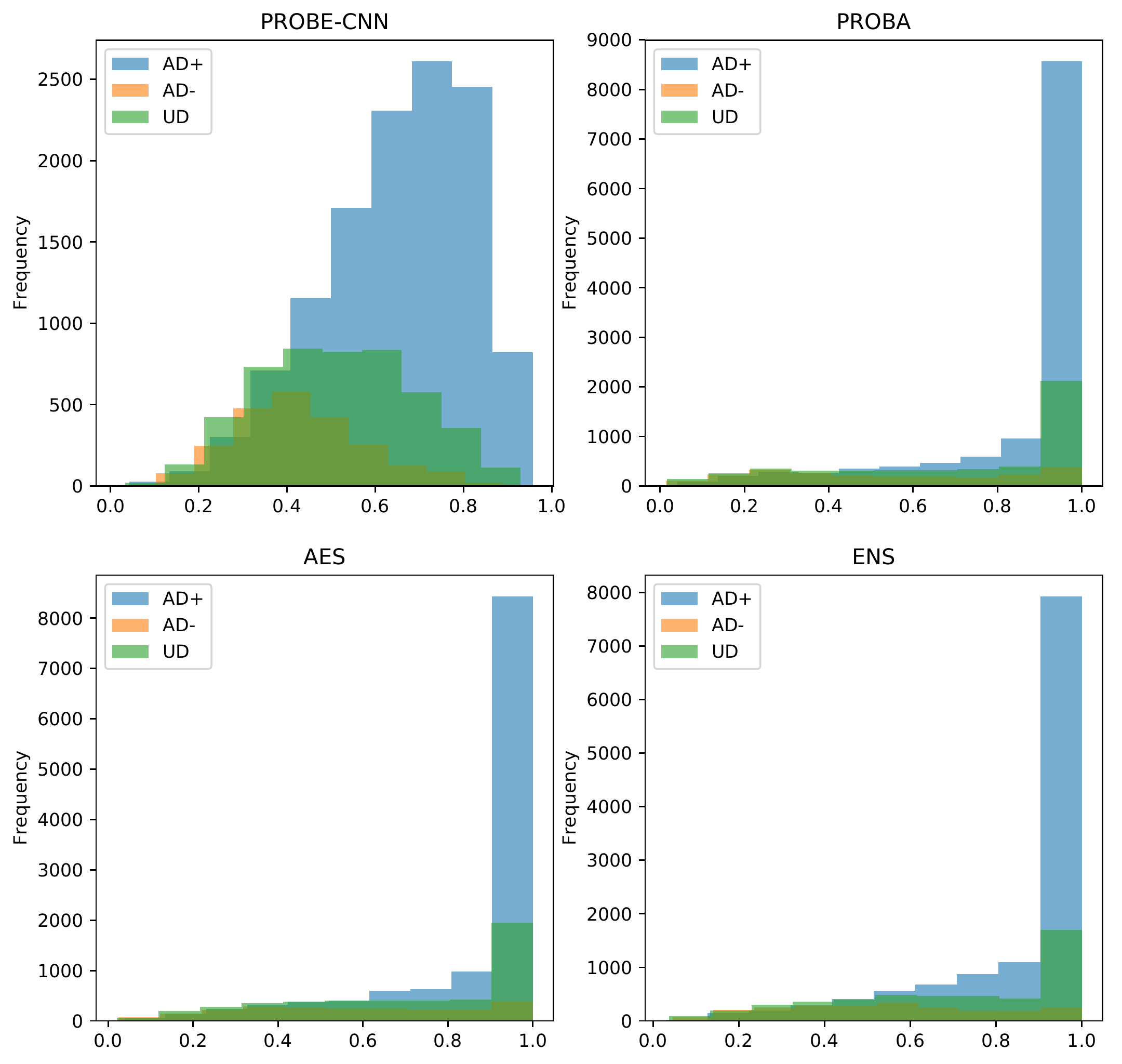}
\caption{Confidence histogram of BIDAF on SogouRC.}
\label{fig:hist}
\end{figure}

\begin{figure*}[t]
\centering
\includegraphics[scale=0.32]{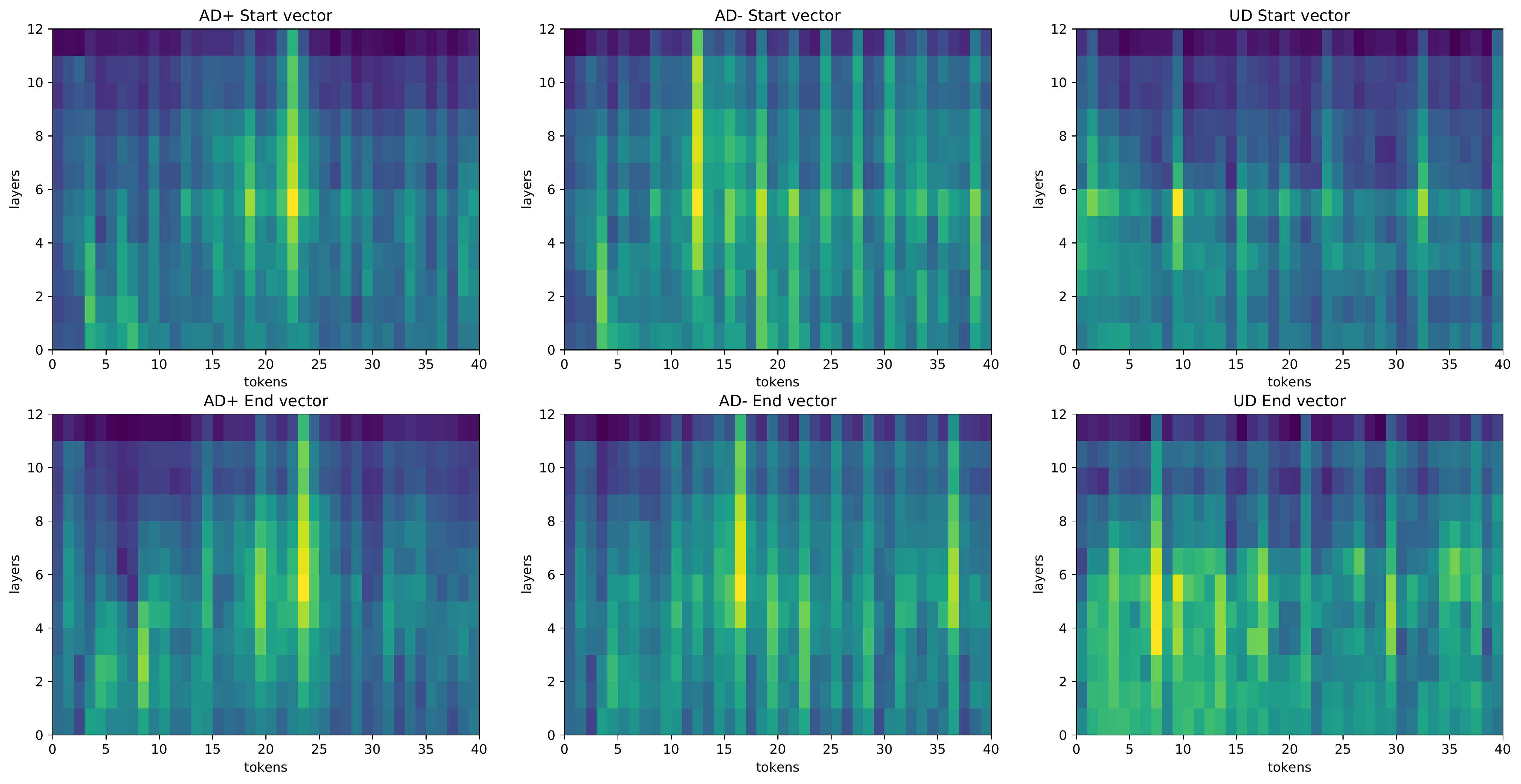}
\caption{Heatmap of the probed signals of three sampled cases from the AD$^+$, AD$^-$ and UD set using BERT on SogouRC .}
\label{fig:heatmap}
\end{figure*}

 \subsection{Detailed Comparison}
Beyond these overall performance comparison, we further take some detailed analysis. Here we draw the RC-curve of the four methods as well as the oracle on the SogouRC dataset using the BIDAF as the basic MRC model, as shown in Figure~\ref{fig:rc_curve}. We can see the ideal RC-curve is the red line from the oracle. When we increase the coverage (by lowering the confidence threshold $\theta$), an ideal risk control framework should first keep the risk at 0 upto certain coverage, and then increase steadily until the coverage reaches 1. Such curve means that the risk control framework can correctly rank the instances based on the predicted confidence score so that there is a rejection region that no risk would happen within it. Among the four methods, we can observe that the RC-curve of our PROBE-CNN method is always under all the other curves and also the most close to the oracle curve, showing that our method can consistently predict better confidence score for all the instances. 

We also depicted the  histogram of predicted confidence scores from the four methods over the three set of instances, i.e., AD$^+$, AD$^-$ and UD, to obtain better understanding. The results are shown in Figure~\ref{fig:hist}. Ideally, we expect a good risk control framework could distinguish these instances clearly and give instances in the AD$^+$ set higher scores than those in the AD$^-$ and UD set. From the results, we can observe that the three baseline methods cannot distinguish them very well as both AD$^+$ and UD have their peak value on the right. Our method can better distinguish these instances in terms of the score distribution, although UD set is still the most difficult one to deal with. This is probably due to that there might be no clear/stable pattern in the UD instances that could be captured by our probes.

Since the UD instances are the most difficult to handle in the risk control framework, some people might wonder whether our method gain the improvement by just better recognizing those UD instances from the rest. Here we conduct a further analysis by comparing all the four risk control methods only on the answerable queries. We use the answerable part in SoguoRC (Named SogouRC Answerable) and SQuAD 1.0 dataset for experiments. The results are shown in Table~\ref{table:answerable}. From the results, we can see that our method can also consistently outperform all the baseline methods in terms of all the evaluation metrics. The results indicate that our PROBE-CNN can well control the risk via distinguishing AD$^+$ from  AD$^-$.

\begin{table}[t]
\caption{Comparison results on the answerable datasets.}
\label{table:answerable}
\small
\centering
\begin{tabular}{cccccccc}
\toprule 
\hline
\multicolumn{2}{c}{\multirow{2}{*}{}} & \multicolumn{3}{c}{SogouRC Answerable} & \multicolumn{3}{c}{SQuAD 1.0} \\ \hline
\multicolumn{2}{c}{}                  & AURC     & ROC     & AP     & AURC     & ROC    & AP    \\ \hline
\multirow{4}{*}{BIDAF}   & PROBA      &         7.38&83.06&55.15        &       8.37&80.29&52.42             \\
                         & AES        &      7.20&83.36&55.92          &        7.96&81.5&54.10 \\
                         & ENS        &      7.06&84.2&56.18       &    7.90&81.63&54.40    \\
                         & Ours       &     6.9&84.71&56.85      &         7.8&81.72&53.88 \\  \hline \midrule
\multirow{4}{*}{BERT}    & PROBA      &          3.22&85.49&47.57       &       5.24&78.22&37.74  \\
                         & AES        &           3.03&86.34&47.56       &      5.10&79.28&44.38\\
                         & ENS        &          2.78&87.50&47.99      &      4.20&83.9&46.90    \\
                         & Ours       &               2.66&88.03&48.93       &    3.63&85.53&49.26    \\  \hline                                 \bottomrule
\end{tabular}

\end{table}

 \subsection{Analysis on the Probe}
 We further conduct experiments to analyze our probe-based qualify model. We first investigate the effect of layer numbers to our PROBE-CNN method. In our original design, we probe each intermediate layer of the MRC model to obtain signals for predictive uncertainty estimation. Here we test the performance if we only probe the last layer $P^{(T)}$ of the MRC model, namely PROBE-CNN$_{P^{(T)}}$. The results are shown in Table~\ref{table:used_layer}. We can see that there is a clear performance drop if we only probe the last layer. The relative decrease of PROBE-CNN$_{P^{T}}$ over the vanilla PROBE-CNN is about 3\% in terms of AURC on the SogouRC dataset. The results demonstrate that the last layer does contain some useful signals, e.g., the distribution of the final $s$ and $e$ scores, but the intermediate layers can bring much richer information for uncertainty estimation.

 Furthermore, We plot the outputs of the probes to gain some intuitive understanding on what the probes have learned. We sample three cases from the AD$^+$, AD$^-$ and UD set using the BERT model respectively and show their probed start and end vector signals in Figure~\ref{fig:heatmap}. Note the vertical axis denotes the layer number and the horizontal axis denotes the word index in the passage, and the brighter the signal, the larger value it is. From the figure we can see that the correct instance (from the AD$^+$ set) shows a distinguishable position consistently from bottom to top. While for the incorrect instances (from the AD$^-$ and UD set), there are often multiple indistinguishable positions with strong signals and the patterns are varying from layer to layer. These probed patterns cope well with our intuition and bring good interpretability to our probe-based qualify model.

\begin{table}[t]
\caption{Comparison results over the probe number.}
\label{table:used_layer}
\begin{tabular}{lllll}
\toprule
\hline
                         &       & AURC  & ROC   & AP    \\ \hline
\multirow{2}{*}{SogouRC} & PROBE-CNN$_{P^{(T)}}$ & 14.60 & 75.12 & 53.1  \\
                         & PROBE-CNN   & 14.16(3.0\%) & 75.46 & 53.8 \\ \hline
\multirow{2}{*}{SQuAD}   & PROBE-CNN$_{P^{(T)}}$ & 24.39 & 68.52 & 55.30 \\
                         & PROBE-CNN   & 23.41(4.0\%) & 69.55 & 56.47\\ \hline
                         \bottomrule
\end{tabular}

\end{table}

\section{Conclusions}
In this paper, we introduced the risk control problem of Web QA by modeling the predictive uncertainty of deep MRC models. We conducted an in-depth analysis of the risk of Web QA. Based on the analysis, we proposed a risk control framework with a novel and general deep qualify model designed based on the abstraction of modern MRC models. We conducted extensive experiments on publicly available benchmark datesets using risk-aware evaluation metrics. The empirical results demonstrate the effectiveness and show the good interpretability of our proposed method. For the future work, we may apply our proposed risk control framework to other IR tasks, e.g., query suggestion.

\begin{acks}
This work was funded by the National Natural Science Foundation of China (NSFC) under Grants No. 61425016, 61722211, 61773362, and 61872338, the Youth Innovation Promotion Association CAS under Grants No. 20144310, and 2016102, the National Key R\&D Program of China under Grants No. 2016QY02D0405, and the Foundation and Frontier Research Key Program of Chongqing Science and Technology Commission (No. cstc2017jcyjBX0059).
\end{acks}


%
\bibliographystyle{ACM-Reference-Format}
\bibliography{sigir19-sigconf}

\end{document}